\documentclass[12pt,epsf]{article}

\usepackage{times}
\usepackage{graphicx}
\usepackage{amsfonts}
\usepackage{amsmath}
\usepackage{amssymb}
\usepackage{amstext}
\usepackage{amsthm}

\usepackage{epsfig}
\usepackage{color}
\psfigdriver{dvips}
\usepackage{latexsym}
\usepackage[matrix,curve]{xypic}
\usepackage{rotating}
\rotdriver{dvips}

\begin{document}

\baselineskip 6mm
\renewcommand{\thefootnote}{\fnsymbol{footnote}}


\newcommand{\nc}{\newcommand}
\newcommand{\rnc}{\renewcommand}


\rnc{\baselinestretch}{1.24}    
\setlength{\jot}{6pt}       
\rnc{\arraystretch}{1.24}   

\makeatletter
\rnc{\theequation}{\thesection.\arabic{equation}}
\@addtoreset{equation}{section}
\makeatother



\nc{\be}{\begin{equation}}

\nc{\ee}{\end{equation}}

\nc{\bea}{\begin{eqnarray}}

\nc{\eea}{\end{eqnarray}}

\nc{\xx}{\nonumber\\}

\nc{\ct}{\cite}

\nc{\la}{\label}

\nc{\eq}[1]{(\ref{#1})}

\nc{\newcaption}[1]{\centerline{\parbox{6in}{\caption{#1}}}}

\nc{\fig}[3]{

\begin{figure}
\centerline{\epsfxsize=#1\epsfbox{#2.eps}}
\newcaption{#3. \label{#2}}
\end{figure}
}


\def\CA{{\cal A}}
\def\CC{{\cal C}}
\def\CD{{\cal D}}
\def\CE{{\cal E}}
\def\CF{{\cal F}}
\def\CG{{\cal G}}
\def\CH{{\cal H}}
\def\CK{{\cal K}}
\def\CL{{\cal L}}
\def\CM{{\cal M}}
\def\CN{{\cal N}}
\def\CO{{\cal O}}
\def\CP{{\cal P}}
\def\CR{{\cal R}}
\def\CS{{\cal S}}
\def\CU{{\cal U}}
\def\CV{{\cal V}}
\def\CW{{\cal W}}
\def\CY{{\cal Y}}
\def\CZ{{\cal Z}}


\def\IB{{\hbox{{\rm I}\kern-.2em\hbox{\rm B}}}}
\def\IC{\,\,{\hbox{{\rm I}\kern-.50em\hbox{\bf C}}}}
\def\ID{{\hbox{{\rm I}\kern-.2em\hbox{\rm D}}}}
\def\IF{{\hbox{{\rm I}\kern-.2em\hbox{\rm F}}}}
\def\IH{{\hbox{{\rm I}\kern-.2em\hbox{\rm H}}}}
\def\IN{{\hbox{{\rm I}\kern-.2em\hbox{\rm N}}}}
\def\IP{{\hbox{{\rm I}\kern-.2em\hbox{\rm P}}}}
\def\IR{{\hbox{{\rm I}\kern-.2em\hbox{\rm R}}}}
\def\IZ{{\hbox{{\rm Z}\kern-.4em\hbox{\rm Z}}}}


\def\a{\alpha}
\def\b{\beta}
\def\d{\delta}
\def\ep{\epsilon}
\def\ga{\gamma}
\def\k{\kappa}
\def\l{\lambda}
\def\s{\sigma}
\def\t{\theta}
\def\w{\omega}
\def\G{\Gamma}


\def\half{\frac{1}{2}}
\def\dint#1#2{\int\limits_{#1}^{#2}}
\def\goto{\rightarrow}
\def\para{\parallel}
\def\brac#1{\langle #1 \rangle}
\def\curl{\nabla\times}
\def\div{\nabla\cdot}
\def\p{\partial}


\def\Tr{{\rm Tr}\,}
\def\det{{\rm det}}


\def\vare{\varepsilon}
\def\zbar{\bar{z}}
\def\wbar{\bar{w}}
\def\what#1{\widehat{#1}}


\def\ad{\dot{a}}
\def\bd{\dot{b}}
\def\cd{\dot{c}}
\def\dd{\dot{d}}
\def\so{SO(4)}
\def\bfr{{\bf R}}
\def\bfc{{\bf C}}
\def\bfz{{\bf Z}}

\begin{titlepage}


\hfill\parbox{3.7cm} {{\tt arXiv:1101.5185}}

\vspace{15mm}

\begin{center}
{\Large \bf  Einstein Manifolds As Yang-Mills Instantons}

\vspace{10mm}

John J. Oh${}^a$\footnote{johnoh@nims.re.kr} and Hyun Seok
Yang${}^b$\footnote{hsyang@sogang.ac.kr}
\\[10mm]

${}^a$ {\sl Division of Computational Sciences in Mathematics, \\
National Institute for Mathematical Sciences, Daejeon 305-340, Korea}

${}^b$ {\sl Institute for the Early Universe, Ewha Womans
University, Seoul 120-750, Korea}

${}^b$ {\sl Center for Quantum Spacetime, Sogang University, Seoul
121-741, Korea}

\end{center}

\thispagestyle{empty}

\vskip1cm


\centerline{\bf ABSTRACT}
\vskip 4mm
\noindent

It is well-known that Einstein gravity can be formulated as a gauge
theory of Lorentz group where spin connections play a role of gauge
fields and Riemann curvature tensors correspond to their field
strengths. One can then pose an interesting question: What is the
Einstein equation from the gauge theory point of view? Or
equivalently, what is the gauge theory object corresponding to
Einstein manifolds? We show that the Einstein equations in four
dimensions are precisely self-duality equations in Yang-Mills gauge
theory and so Einstein manifolds correspond to Yang-Mills instantons
in $SO(4) = SU(2)_L \times SU(2)_R$ gauge theory. Specifically, we
prove that any Einstein manifold with or without a cosmological
constant always arises as the sum of $SU(2)_L$ instantons and
$SU(2)_R$ anti-instantons. This result explains why an Einstein
manifold must be stable because two kinds of instantons belong to
different gauge groups, instantons in $SU(2)_L$ and anti-instantons
in $SU(2)_R$, and so they cannot decay into a vacuum. We further
illuminate the stability of Einstein manifolds by showing that they
carry nontrivial topological invariants.
\\



\vspace{1cm}

\today

\end{titlepage}

\renewcommand{\thefootnote}{\arabic{footnote}}
\setcounter{footnote}{0}

\section{Introduction}

It seems that the essence of the method of physics is inseparably
connected with the problem of interplay between local and global
aspects of the world's structure, as saliently exemplified in the
index theorem of Dirac operators. Although Einstein field equations,
being differential equations, are defined locally, they have to
determine the structure of spacetime manifold on which they act,
when a boundary condition for the differential equations is properly
taken into account. Therefore, the local character of the Einstein
equations would be intimately connected with the global topological
structure of the underlying manifold \ct{big-book}. The purpose of
this letter is to explore how the topology of spacetime fabric is
encoded into the local structure of Riemannian metrics using the
gauge theory formulation of Euclidean gravity \ct{opy}. It turns out
that the gauge theory formulation of gravity directly reveals the
topological aspects of Einstein manifolds.

The physics on a curved spacetime becomes more transparent when
expressed in a locally inertial frame and it is even indispensable
when one want to couple spinors to gravity since spinors form a
representation of $SO(4)$ rather than $GL(4, \mathbb{R})$. In this
tetrad formalism, a Riemannian metric on spacetime manifold $M$ is
replaced by a local basis for the tangent bundle $TM$, which is
orthonormal tangent vectors $E_A \, (A=1,\cdots, 4)$ on $M$. But, in
any vector space, there is a freedom for the choice of basis and
physical observables are independent of the arbitrary choice of a
tetrad. As in any other gauge theory with local gauge invariance, to
achieve local Lorentz invariance requires introducing a gauge field
${\omega^A}_B$ of the Lorentz group $SO(4)$. The gauge field of the
local Lorentz group is called the spin connection. In the end,
four-dimensional Einstein gravity can be formulated as a gauge
theory of $SO(4)$ Lorentz group where spin connections play a role
of gauge fields and Riemann curvature tensors correspond to their
field strengths.

One can then pose an interesting question: What is the Einstein
equation from the gauge theory point of view? Or equivalently, what
is the gauge theory object corresponding to Einstein manifolds?

In order to answer to the above question, it will be important to
notice the following mystic features \ct{besse,4man-book,4man-book2} existent
only in the four dimensional space. Among the group of isometries of
$d$-dimensional Euclidean space $\mathbb{R}^d$, the Lie group
$SO(4)= Spin(4)/\mathbb{Z}_2$ for $d \ge 3$ is the only non-simple
Lorentz group and one can define a self-dual two-form only for $d =
4$. We will answer to the above question by noting such a plain fact
that the Lorentz group $Spin(4)$ is isomorphic to $SU(2)_L \times
SU(2)_R$ and the Riemann curvature tensors ${R^A}_B = d{\omega^A}_B
+ {\omega^A}_C \wedge {\omega^C}_B$ are $Spin(4)$-valued two-forms.
One can thus apply two kinds of decomposition to spin connections
and curvature tensors. The first decomposition is that the spin
connections ${\omega^A}_B$ can be split into a pair of $SU(2)_L$ and
$SU(2)_R$ gauge fields according to the splitting of the Lie algebra
$SO(4) = SU(2)_L \oplus SU(2)_R$. Accordingly the Riemann curvature
tensors ${R^A}_B$ will also be decomposed into a pair of $SU(2)_L$
and $SU(2)_R$ curvature two-forms. The second decomposition is that,
in four dimensions, the six-dimensional vector space $\Lambda^2
T^*M$ of two-forms splits canonically into the sum of
three-dimensional vector spaces of self-dual and anti-self-dual two
forms, i.e., $\Lambda^2 T^*M = \Lambda^2_+ \oplus
\Lambda^2_-$ \ct{besse,4man-book,4man-book2}. Therefore the Riemann curvature
tensors ${R^A}_B$ will be split into a pair of self-dual and
anti-self-dual two-forms. One can eventually combine these two
decompositions.

Interestingly, the chiral splitting of $SO(4) = SU(2)_L \oplus
SU(2)_R$ and the Hodge decomposition $\Lambda^2 T^*M =
\Lambda^2_+ \oplus \Lambda^2_-$ of two-forms are deeply correlated with each other
due to the isomorphism between the Clifford algebra $\mathbb{C}l(4)$
in four-dimensions and the exterior algebra $\Lambda^* M =
\bigoplus_{k=0}^4 \Lambda^k T^* M$ over a four-dimensional
Riemannian manifold $M$ \ct{spin-book}. In particular, the Clifford
map implies that the $SO(4)$ Lorentz generators $J^{AB} =
\frac{1}{4}[\Gamma^A, \Gamma^B]$ in $\mathbb{C}l(4)$ have one-to-one correspondence
with the space $\Lambda^2 T^*M$ of two-forms in $\Lambda^* M$. Since
the spinor representation in even dimensions is reducible and its
irreducible representations are defined by the chiral
representations whose Lorentz generators are given by $J^{AB}_\pm
\equiv \frac{1}{2}(1 \pm \Gamma^{5})J^{AB}$. Then the splitting of the Lie algebra
$SO(4) = SU(2)_L \oplus SU(2)_R$ can be specified by the chiral
generators $J^{AB}_\pm$ as $J^{AB}_+ \in SU(2)_L$ and $J^{AB}_- \in
SU(2)_R$ and the chiral splitting is precisely isomorphic to the
decomposition $\Lambda^2 T^*M = \Lambda^2_+ \oplus \Lambda^2_-$ of
two-forms on an orientable four-manifold. It would be worthwhile to remark that
these two decompositions actually occupy a central position in the
Donaldson's theory of four-manifolds \ct{4man-book}.

In this paper we will systematically apply the gauge theory
formulation of Einstein gravity to four-dimensional Riemannian
manifolds and consolidate the chiral splitting of Lorentz group and
the Hodge decomposition of two-forms into the gauge theory
formulation. A remarkable result, stated as a lemma in Sec. 2, comes
out which sheds light on why the action of Einstein gravity is
linear in curvature tensors in contrast to the quadratic action of
Yang-Mills theory in spite of a close similarity to gauge theory. It
directly reveals the topological aspects of Einstein manifolds. It
may be emphasized that our result is valid for general Einstein
manifolds with a spin structure and thus generalizes the result for
half-flat manifolds (the so-called gravitational instantons) which
has been well-established as presented in a renowned review
\cite{egh-report} and a textbook \ct{besse}. Our result also directs a new understanding
to the Einstein equations.

The paper is organized as follows.

In Sec. 2, we apply the gauge theory formulation to four-dimensional
Riemannian manifolds. We show that the Einstein equations in four
dimensions are precisely self-duality equations in Yang-Mills gauge
theory and so Einstein manifolds correspond to Yang-Mills instantons
in $SO(4) = SU(2)_L \times SU(2)_R/\mathbb{Z}_2$ gauge theory.
Specifically, we will prove a lemma to state that any Einstein
manifold with or without a cosmological constant always arises as
the sum of $SU(2)_L$ instantons and $SU(2)_R$ anti-instantons. This
result explains why an Einstein manifold must be stable against
small perturbations because two kinds of instantons belong to
different gauge groups, instantons in $SU(2)_L$ and anti-instantons
in $SU(2)_R$, and so they cannot decay into a vacuum.

In Sec. 3, we further illuminate the stability of Einstein
manifolds by showing that they carry nontrivial topological
invariants.

In Sec. 4, we will consider a coupling with gauge fields to
understand how matter fields affect the structure of a vacuum
Einstein manifold.

Finally, in Sec. 5, we discuss how our approach can be applied to
get a new solution of Yang-Mills instantons from a given Einstein
manifold \ct{opy}. An open problem such as the generalization to
other dimensions, e.g., to three and five dimensions will be briefly
discussed.

\section{Einstein manifolds and Yang-Mills instantons}

Four-dimensional Euclidean gravity can be formulated as a gauge
theory using the language of $SO(4)$ gauge theory where the spin
connections ${\omega^A}_B$ are gauge fields with respect to $SO(4)$
rotations. We will follow Ref. \ct{opy} for the gauge theory
formulation of Einstein gravity and also adopt the index notations
in Ref. \ct{opy} except that we further distinguish the two kinds of Lie
algebra indices with $a = 1,2,3$ and $\dot{a}=1,2,3$ for $SU(2)_L$
and $SU(2)_R$, respectively, in $Spin(4) = SU(2)_L \times SU(2)_R$
Lorentz group. In particular, the identities for the 't Hoof symbols
(Eqs. (3.13)-(3.19) in Ref. \ct{opy}) will be extensively used in this work.

Suppose that $M$ is an oriented four-manifold with a spin structure,
i.e., the second Stiefel-Whitney class $w_2 \in H^2(M,\mathbb{Z}_2)$
identically vanishes. The Hodge $*$-operation defines an
automorphism of the vector space $\Lambda^2 T^*M$ of two-forms with
the decomposition
\begin{equation}\label{2-form-dec}
    \Lambda^2 T^*M = \Lambda^+_3 \oplus \Lambda^-_3
\end{equation}
where $\Lambda^\pm_3 \equiv P_\pm \Lambda^2 T^*M$ and $P_\pm =
\frac{1}{2}(1 \pm *)$. The Hodge decomposition \eq{2-form-dec} can be
harmoniously incorporated with the Lie algebra isomorphism $SO(4) =
SU(2)_L \oplus SU(2)_R$ according to the Clifford isomorphism
$\mathbb{C}l(4) \cong \Lambda^* M$. In this respect, the 't Hooft
symbols $\eta^a_{AB}$ and $\overline{\eta}^{\dot{a}}_{AB}$ take a
superb mission consolidating the Hodge decomposition
\eq{2-form-dec} and the Lie algebra isomorphism $SO(4) = SU(2)_L
\oplus SU(2)_R$, which intertwine the group structure carried by
the Lie algebra indices $a =1,2,3 \in SU(2)_L$ and $\dot{a}=1,2,3
\in SU(2)_R$ with the spacetime structure of two-form indices $A,B$.

Since the spin connections $\omega_{AB} = \omega_{MAB} dx^M$ are
gauge fields taking values in $SO(4)$ Lie algebra, first let us
apply the Lie algebra decomposition $SO(4) = SU(2)_L \oplus SU(2)_R$
to them. This can explicitly be realized by considering the
following splitting of spin connections \ct{opy}
\begin{equation} \label{spin-sd-asd}
\omega_{MAB} \equiv A_M^{(+)a} \eta^a_{AB} + A_M^{(-)\dot{a}}
\overline{\eta}^{\dot{a}}_{AB}
\end{equation}
where $A^{(+)a} = A_M^{(+)a} dx^M$ and $A^{(-)\dot{a}} =
A_M^{(-)\dot{a}} dx^M$ are $SU(2)_L$ and $SU(2)_R$ gauge fields,
respectively. The Riemann curvature tensors $R_{AB} =
\frac{1}{2} R_{MNAB} dx^M \wedge dx^N$ then take a similar decomposition
\begin{equation}\label{riemann-su2}
 R_{MNAB} \equiv F^{(+)a}_{MN}\eta^a_{AB} + F^{(-)\dot{a}}_{MN}
 \overline{\eta}^{\dot{a}}_{AB},
\end{equation}
where
\begin{equation} \label{gi-curvature}
F_{MN}^{(\pm)} = \partial_M A_N^{(\pm)} - \partial_N A_M^{(\pm)} +
[A_M^{(\pm)}, A_N^{(\pm)}]
\end{equation}
are field strengths of $SU(2)_L$ and $SU(2)_R$ gauge fields in Eq.
\eq{spin-sd-asd}. Now we will give an answer to the question raised
before.

$\textbf{Lemma}.$\; The Riemann curvature two-forms $R_{AB} =
\frac{1}{2} R_{MNAB} dx^M \wedge dx^N$ are $SO(4)$-valued field strengths of the spin connections
in Eq. \eq{spin-sd-asd} from the gauge theory point of view and thus
can be decomposed into a pair of $SU(2)_L$ and $SU(2)_R$ field
strengths. With the decomposition \eq{riemann-su2}, the Einstein
equations
\begin{equation} \la{einstein-equation}
R_{AB} - \frac{1}{2}\delta_{AB}R + \delta_{AB}\Lambda=0
\end{equation}
for a Riemannian manifold $M$ are equivalent to the self-duality
equations
\begin{equation} \la{ym-instanton-eq}
F^{(\pm)}_{AB} = \pm
\frac{1}{2}{\varepsilon_{AB}}^{CD}F^{(\pm)}_{CD}
\end{equation}
of Yang-Mills instantons where $F^{(+)a}_{AB}\eta^a_{AB} =
F^{(-)\dot{a}}_{AB}\overline{\eta}^{\dot{a}}_{AB}=2\Lambda$.

{\it Proof}. \; According to the Lie algebra splitting $SO(4) =
SU(2)_L \oplus SU(2)_R$, the Riemann curvature tensors $R_{AB} =
\frac{1}{2} R_{MNAB} dx^M \wedge dx^N$ in Eq. \eq{riemann-su2} have been decomposed
into a pair of  $SU(2)_L$ and $SU(2)_R$ field strengths defined by
$F^{(+)a} = \frac{1}{2} F^{(+)a}_{MN} dx^M \wedge dx^N$ and
$F^{(-)\dot{a}}= \frac{1}{2} F^{(-)\dot{a}}_{MN} dx^M \wedge dx^N$,
respectively. Because $F^{(\pm)}$ are curvature two-forms in gauge
theory, we can apply the Hodge decomposition \eq{2-form-dec} to the $SU(2)$ field strengths $F^{(\pm)}_{AB}
\equiv E_A^M E_B^N F^{(\pm)}_{MN}$ as follows
\begin{eqnarray} \la{su2-f+}
&& F^{(+)a}_{AB} \equiv f^{ab}_{(++)}\eta^b_{AB} +
f^{a\dot{b}}_{(+-)} \overline{\eta}^{\dot{b}}_{AB}, \la{dec-su2l} \\
\la{su2-f-}
&& F^{(-)\dot{a}}_{AB} \equiv f^{\dot{a}b}_{(-+)}\eta^b_{AB} +
f^{\dot{a}\dot{b}}_{(--)} \overline{\eta}^{\dot{b}}_{AB}.
\la{dec-su2r}
\end{eqnarray}
Using the above result, we get the general decomposition of the
Riemann curvature tensor given by
\begin{equation} \la{dec-riemann}
 R_{ABCD}  = f^{ab}_{(++)}\eta^a_{AB} \eta^b_{CD}
 + f^{a\dot{b}}_{(+-)} \eta^a_{AB} \overline{\eta}^{\dot{b}}_{CD}
 + f^{\dot{a}b}_{(-+)} \overline{\eta}^{\dot{a}}_{AB} \eta^b_{CD}
 + f^{\dot{a}\dot{b}}_{(--)} \overline{\eta}^{\dot{a}}_{AB} \overline{\eta}^{\dot{b}}_{CD}.
\end{equation}
Note that the curvature tensor has the symmetry property
$R_{ABCD}=R_{CDAB}$ from which one can get the following relations
between the coefficients in the expansion \eq{dec-riemann}:
\begin{equation}\label{symm-coeff}
f^{ab}_{(++)} = f^{ba}_{(++)}, \quad f^{\dot{a}\dot{b}}_{(--)} =
f^{\dot{b}\dot{a}}_{(--)},
\quad f^{a\dot{b}}_{(+-)} = f^{\dot{b}a}_{(-+)}.
\end{equation}
The first Bianchi identity $\varepsilon^{ACDE}R_{BCDE}=0$ (from which the symmetry property
$R_{ABCD}=R_{CDAB}$ is actually deduced) further constrains the coefficients
\begin{equation}\label{bian-coeff}
  f^{ab}_{(++)}\delta^{ab} = f^{\dot{a}\dot{b}}_{(--)}\delta^{\dot{a}\dot{b}}.
\end{equation}
Hence the Riemann curvature tensor in Eq. \eq{dec-riemann} has $20 =
(6+6-1) + 9$ independent components, as is well-known \ct{big-book}.

The above results can be applied to the Ricci tensor $R_{AB}\equiv
R_{ACBC}$ and the Ricci scalar $R\equiv R_{AA}$ to yield
\begin{eqnarray}  \la{dec-ricci}
 R_{AB} &=& \big(f^{ab}_{(++)} \delta^{ab}
 + f^{\dot{a}\dot{b}}_{(--)} \delta^{\dot{a}\dot{b}}\big) \delta_{AB}
 + 2f^{a\dot{a}}_{(+-)}\eta^a_{AC} \overline{\eta}^{\dot{a}}_{BC}, \\
 \la{dec-scalar}
 R &=& 4 \big(f^{ab}_{(++)} \delta^{ab}
 + f^{\dot{a}\dot{b}}_{(--)} \delta^{\dot{a}\dot{b}}\big),
\end{eqnarray}
where a symmetric expression was taken in spite of the relation
\eq{bian-coeff}. After all, the Einstein tensor $G_{AB} \equiv R_{AB} -
\frac{1}{2} R \delta_{AB}$ has 10 independent components given by
\begin{equation}  \la{dec-einstein}
 G_{AB} =  2f^{a\dot{a}}_{(+-)}\eta^a_{AC} \overline{\eta}^{\dot{a}}_{BC}
 - 2 f^{ab}_{(++)} \delta^{ab} \delta_{AB}.
\end{equation}

Note that the Einstein equation \eq{einstein-equation} can be recast
in the form $R_{AB} = \Lambda \delta_{AB}$ where $\Lambda$ is a
cosmological constant. Therefore the Einstein condition can easily
be read off from Eq. \eq{dec-ricci} and the result is given by
\begin{equation}\label{cond-einstein}
  f^{ab}_{(++)} \delta^{ab} = f^{\dot{a}\dot{b}}_{(--)} \delta^{\dot{a}\dot{b}}
  = \frac{\Lambda}{2}, \qquad
  f^{a\dot{a}}_{(+-)} = 0.
\end{equation}
Therefore, the curvature tensor for an Einstein manifold reduces to
 \begin{eqnarray} \label{einstein-mfd}
 R_{ABCD} &=& F^{(+)a}_{AB}\eta^a_{CD}
 + F^{(-)\dot{a}}_{AB} \overline{\eta}^{\dot{a}}_{CD} \xx
 &=& f^{ab}_{(++)} \eta^a_{AB} \eta^b_{CD}
 + f^{\dot{a}\dot{b}}_{(--)} \overline{\eta}^{\dot{a}}_{AB}
 \overline{\eta}^{\dot{b}}_{CD}
\end{eqnarray}
with the coefficients satisfying \eq{cond-einstein}. Eq.
\eq{einstein-mfd} immediately shows that $F^{(\pm)}_{AB}$ are
$SU(2)$ field strengths obeying the self-duality equations in Eq.
\eq{ym-instanton-eq}.

And one can verify that the converse is true too: If the Riemann
curvature tensor is given by Eq. \eq{einstein-mfd} and so satisfies
the self-duality equations \eq{ym-instanton-eq}, the Einstein
equations \eq{einstein-equation} are automatically satisfied with
$2\Lambda = F^{(+)a}_{AB}\eta^a_{AB} =
F^{(-)\dot{a}}_{AB}\overline{\eta}^{\dot{a}}_{AB}$. This completes
the proof of the Lemma. $\qquad \Box$

Let us consider special classes of Einstein manifolds to illustrate
how they easily comply with our general result. First, for
gravitational instantons satisfying
\begin{equation}\label{g-instanton}
R_{EFAB} = \frac{1}{2} {\varepsilon_{AB}}^{CD} R_{EFCD},
\end{equation}
we get the curvature tensor \ct{opy}
\begin{equation}\label{gi-riemann}
R_{ABCD}  = F^{(+)a}_{AB}\eta^a_{CD} = f^{ab}_{(++)}\eta^a_{AB}
\eta^b_{CD}
\end{equation}
with $f^{ab}_{(++)}\delta^{ab} = 0$. Therefore the gravitational
instanton is half-flat, i.e. $F^{(-)\dot{a}}_{AB} = 0$ and
Ricci-flat, i.e. $f^{ab}_{(++)}\delta^{ab} = 0$. Similarly, for
gravitational anti-instantons satisfying
\begin{equation}\label{g-ainstanton}
R_{EFAB} = - \frac{1}{2} {\varepsilon_{AB}}^{CD} R_{EFCD},
\end{equation}
the curvature tensor is given by
\begin{equation}\label{agi-riemann}
R_{ABCD}  = F^{(-)\dot{a}}_{AB} \overline{\eta}^{\dot{a}}_{CD} =
f^{\dot{a}\dot{b}}_{(--)} \overline{\eta}^{\dot{a}}_{AB}
\overline{\eta}^{\dot{b}}_{CD}
\end{equation}
with $f^{\dot{a}\dot{b}}_{(--)} \delta^{\dot{a}\dot{b}} = 0$.

From the results \eq{gi-riemann} and \eq{agi-riemann}, one can
easily see that gravitational instantons are $SU(2)$ Yang-Mills
instantons in the sense that they satisfy the self-duality equations
\eq{ym-instanton-eq}. Actually this result is not new but has been well understood as presented in
well-known reviews \ct{besse,egh-report}. Anyway it is interesting
to notice that the solution of $F^{(\pm)}_{AB} = 0$ corresponds to a
Ricci-flat, K\"ahler manifold and so it describes a Calabi-Yau
2-fold with $SU(2)$ holonomy. In other words, hyper-K\"ahler
manifolds can be recast into the self-dual connections defined by
Yang-Mills instantons \ct{loy1}. Indeed, one can easily show that
self-dual connections satisfying the half-flat condition
$F^{(-)\dot{a}}_{AB} = 0$ admit the triple of K\"ahler forms defined
by
\be \la{hyper-kahler}
J^a_+ = \frac{1}{2} \eta^a_{AB} E^A \wedge E^B, \qquad a=1,2,3
\ee
which are all closed, i.e., $dJ^a_+ = 0$. Similarly, it is easy to
show that  anti-self-dual connections satisfying the half-flat
condition $F^{(+)a}_{AB} = 0$ guarantee the existence of the triple
of K\"ahler forms defined by
\be \la{anti-hyper-kahler}
J^{\dot{a}}_- = \frac{1}{2} \overline{\eta}^{\dot{a}}_{AB} E^A
\wedge E^B, \qquad \dot{a}=1,2,3
\ee
which are also closed 2-forms, $dJ^{\dot{a}}_- = 0$.

For a Ricci-flat manifold obeying $R_{AB}=0$, we get the condition
from Eq. \eq{dec-ricci}
\begin{equation}\label{cond-ricci}
  f^{ab}_{(++)} \delta^{ab} = f^{\dot{a}\dot{b}}_{(--)} \delta^{\dot{a}\dot{b}} = 0, \qquad
  f^{a\dot{a}}_{(+-)} = 0
\end{equation}
and so the following decomposition
\begin{eqnarray} \label{ricci-flat}
 R_{ABCD} &=& F^{(+)a}_{AB}\eta^a_{CD} + F^{(-)\dot{a}}_{AB} \overline{\eta}^{\dot{a}}_{CD} \xx
 &=& f^{ab}_{(++)}\eta^a_{AB} \eta^b_{CD}
 + f^{\dot{a}\dot{b}}_{(--)} \overline{\eta}^{\dot{a}}_{AB} \overline{\eta}^{\dot{b}}_{CD}
\end{eqnarray}
with the traceless coefficients satisfying Eq. \eq{cond-ricci}. Thus
Eq. \eq{ricci-flat} is a particular case with $\Lambda =0$ of the
general result \eq{einstein-mfd}.

The decomposition \eq{ricci-flat} of Riemann curvature tensors for a
Ricci-flat manifold is consistent with the double-dual condition
\be \la{double-dual}
{\varepsilon_{AB}}^{A'B'} R_{A'B'CD} =  R_{ABC'D'}
{\varepsilon^{C'D'}}_{CD}
\ee
first introduced by Charap and Duff \ct{charap-duff,charap-duff2}. One can easily
check that the curvature tensor in Eq. \eq{ricci-flat} obeys the
double-dual condition \eq{double-dual} by using the self-duality
relations for 't Hooft symbols:
\be \la{sd-tsymbol}
\eta^a_{AB} = \frac{1}{2} {\varepsilon_{AB}}^{CD}
\eta^a_{CD}, \qquad  \overline{\eta}^{\dot{a}}_{AB} = - \frac{1}{2} {\varepsilon_{AB}}^{CD}
\overline{\eta}^{\dot{a}}_{CD}.
\ee
It was noted in Refs. \ct{charap-duff,charap-duff2} that Ricci-flat spaces satisfy the
condition \eq{double-dual} whose solution can be used to construct
$SU(2)$ self-dual connections (Yang-Mills instantons) on a
Ricci-flat manifold. For example, Euclidean Schwarzschild black-hole
is a Ricci-flat manifold \ct{hawking-inst} and so the self-dual part
of its spin connections can be implemented to find a Yang-Mills
instanton on the black-hole geometry. However, our result
\eq{einstein-mfd} shows that not only a Ricci-flat manifold but also
a general Einstein manifold obeys the double-dual condition
\eq{double-dual} and the Einstein manifold can always be split into
$SU(2)_L$ instantons and $SU(2)_R$ anti-instantons. Thus we have
generalized the result in Refs. \ct{charap-duff,charap-duff2} to Einstein
manifolds, which has not been addressed so far.

One can draw a very interesting implication from the lemma we have
proven. The $SU(2)$ field strengths in Eq. \eq{riemann-su2} are
given by
\begin{equation}\label{su2-curvature}
    F^{(\pm)} =  dA^{(\pm)} + A^{(\pm)} \wedge A^{(\pm)}.
\end{equation}
The integrability condition, namely, the Bianchi identity, then
reads as
\begin{equation}\label{su2-bianchi}
D^{(\pm)} F^{(\pm)} \equiv dF^{(\pm)} + A^{(\pm)} \wedge F^{(\pm)} -
F^{(\pm)} \wedge A^{(\pm)} = 0.
\end{equation}
Therefore the self-duality equation \eq{ym-instanton-eq} immediately
leads to the remarkable result that any Einstein manifold
automatically satisfies the Yang-Mills equations of motion, i.e.,
\begin{equation}\label{ym-eq}
D^{(\pm)} *F^{(\pm)} = \pm D^{(\pm)} F^{(\pm)} = 0 \;\;
\Leftrightarrow \;\; D *F = D^{(+)} *F^{(+)} +  D^{(-)} *F^{(-)} = 0
\end{equation}
where $*F$ means the Hodge $*$-operation on a two-form $F$. After
all, our lemma sheds light on why the action of Einstein gravity is
linear in curvature tensors contrary to the Yang-Mills action being
quadratic in curvatures. If the action of Einstein gravity were
quadratic in curvature tensors, four-manifolds obeying the equations
of motion would not necessarily be given by $SU(2)$ Yang-Mills
instantons and the four-manifold could be unstable in general as is
well-known from gauge theory. Furthermore our lemma poses an
intriguing issue about how to quantize an Einstein manifold, which
will be discussed in the last section.

The ``trace-free part" of the Riemann curvature tensor is called the
Weyl tensor \ct{big-book} defined by
\begin{equation}\label{weyl}
W_{ABCD} = R_{ABCD} - \frac{1}{2} \big(\delta_{AC} R_{BD} -
\delta_{AD} R_{BC} - \delta_{BC} R_{AD}  + \delta_{BD} R_{AC} \big)
+ \frac{1}{6} (\delta_{AC} \delta_{BD}- \delta_{AD} \delta_{BC}) R.
\end{equation}
The Weyl tensor shares all the symmetry structures of the curvature
tensor and all its traces with the metric are zero. Therefore, one
can introduce a similar decomposition for the Weyl tensor
\begin{equation} \la{dec-weyl}
 W_{ABCD}  \equiv g^{ab}_{(++)}\eta^a_{AB} \eta^b_{CD}
 + g^{a\dot{a}}_{(+-)} \eta^a_{AB} \overline{\eta}^{\dot{a}}_{CD}
 + g^{\dot{a}a}_{(-+)} \overline{\eta}^{\dot{a}}_{AB} \eta^a_{CD}
 + g^{\dot{a}\dot{b}}_{(--)} \overline{\eta}^{\dot{a}}_{AB} \overline{\eta}^{\dot{b}}_{CD}.
\end{equation}
The symmetry property of the coefficients in the expansion
\eq{dec-weyl} is the same as Eq. \eq{symm-coeff} and the traceless
condition, i.e. $W_{AB} \equiv W_{ACBC} = 0$, leads to the
constraint for the coefficients:
\begin{equation}\label{cond-weyl}
  g^{ab}_{(++)} \delta^{ab} = g^{\dot{a}\dot{b}}_{(--)} \delta^{\dot{a}\dot{b}} = 0, \qquad
  g^{a\dot{a}}_{(+-)} = g^{\dot{a}a}_{(-+)}= 0.
\end{equation}
Hence the $SO(4)$-decomposition for the Weyl tensor is finally given
by
\begin{equation} \la{final-weyl}
 W_{ABCD}  = g^{ab}_{(++)}\eta^a_{AB} \eta^b_{CD}
 + g^{\dot{a}\dot{b}}_{(--)} \overline{\eta}^{\dot{a}}_{AB} \overline{\eta}^{\dot{b}}_{CD}
\end{equation}
with the coefficients satisfying \eq{cond-weyl}. One can see that
the Weyl tensor has only $10 = 5+5$ independent components.

It is straightforward to determine the expansion coefficients
$g^{ab}_{(++)} = \frac{1}{16} \eta^a_{AB} \eta^b_{CD}  W_{ABCD}$ and
$g^{\dot{a}\dot{b}}_{(--)} = \frac{1}{16}
\overline{\eta}^{\dot{a}}_{AB} \overline{\eta}^{\dot{b}}_{CD}  W_{ABCD}$ in Eq. \eq{final-weyl}
in terms of the coefficients in curvature tensors by substituting
the results \eq{dec-riemann} and \eq{dec-ricci} into Eq. \eq{weyl}:
\begin{equation} \la{weyl-coeff}
g^{ab}_{(++)} =  f^{ab}_{(++)} - \frac{1}{3} \delta^{ab}
f^{cd}_{(++)} \delta^{cd}, \qquad g^{\dot{a}\dot{b}}_{(--)}  =
f^{\dot{a}\dot{b}}_{(--)} - \frac{1}{3} \delta^{\dot{a}\dot{b}}
f^{\dot{c}\dot{d}}_{(--)} \delta^{\dot{c}\dot{d}}.
\end{equation}
Then Eq. \eq{final-weyl} can be written as follows
\begin{equation} \la{f-weyl}
W_{ABCD}  = f^{ab}_{(++)}\eta^a_{AB} \eta^b_{CD} +
f^{\dot{a}\dot{b}}_{(--)} \overline{\eta}^{\dot{a}}_{AB}
\overline{\eta}^{\dot{b}}_{CD} - \frac{1}{3}
\big(f^{ab}_{(++)} \delta^{ab} + f^{\dot{a}\dot{b}}_{(--)}
\delta^{\dot{a}\dot{b}} \big) (\delta_{AC}
\delta_{BD}- \delta_{AD} \delta_{BC}).
\end{equation}
Combining the results in Eqs. \eq{dec-riemann} and \eq{f-weyl} gives
us the well-known decomposition of the curvature tensor $R$ into
irreducible components \ct{ahs-math}, schematically given by
\begin{equation} \label{ahs-dec}
R = \left(
    \begin{array}{cc}
      W^+ + \frac{1}{12} s &  B  \\
      B^T &  W^- + \frac{1}{12} s \\
    \end{array}
  \right),
\end{equation}
where $s$ is the scalar curvature, $B$ is the traceless Ricci
tensor, and $W^\pm$ are the (anti-)self-dual Weyl tensors. One can
check that imposing the Einstein equations $R_{AB} = \Lambda
\delta_{AB}$ upon the decomposition \eq{ahs-dec} leads to the
condition, $s = 4
\Lambda$ and $B = B^T = 0$, consistent with the Lemma in Sec. 2.
Therefore our Lemma is not completely new but rather well-known in
Riemannian geometry. Nevertheless the statement of the Lemma in
terms of the explicit decomposition \eq{dec-riemann} of Riemann
curvature tensors will benefit by several applications such as the
calculation of topological invariants, the generalization to matter
couplings and the construction of Yang-Mills instantons from
Einstein manifolds, as will be addressed in the following sections.

One can consider the self-duality equation for Weyl tensors defined
by $W_{EFAB} = \pm \frac{1}{2} {\varepsilon_{AB}}^{CD} W_{EFCD}$
\ct{egh-report}. An Einstein manifold is conformally self-dual if
$g^{\dot{a}\dot{b}}_{(--)}=0$ and conformally anti-self-dual if
$g^{ab}_{(++)} = 0$ in Eq. \eq{final-weyl}. Note that the Weyl
instanton (a conformally self-dual manifold) can also be regarded as
a Yang-Mills instanton and $\mathbb{C}P^2$ is a well-known example
\ct{gibb-pope}.

But there is a subtle point for instantons with a non-zero
cosmological constant. One can see from the condition
\eq{cond-einstein} that $SU(2)$ field strengths in Eq.
\eq{einstein-mfd} do not decay to zero at an asymptotic region. It
is not a problem for the case with $\Lambda > 0$, e.g. a de Sitter
space, because these spaces such as $\mathbb{S}^4$ and
$\mathbb{C}P^2$ are all compact \ct{gibb-hawk-cmp}. So the
corresponding Yang-Mills action can be finite even with the
asymptotic condition \eq{cond-einstein}. A trouble arises in the
case with $\Lambda < 0$, e.g. an anti-de Sitter space, because the
gravitational action will diverge for noncompact geometries. To
define a finite action for noncompact geometries, we may choose a
reference background such that the physical (regularized) action of
the reference background is defined to be zero as the ground state
\ct{hawk-horo,ads-inst,ads-inst2} by subtracting an infinite contribution from the
background solution. From the gauge theory point of view, this
regularization can be realized \ct{future} by expanding $SO(4)$
gauge fields $A_M^{(\pm)}$ around a classical background field
$B_M^{(\pm)}$ (a.k.a., the background field method).

\section{Topological invariants and stability of Einstein manifolds}

Since Einstein manifolds carry a topological information in the form
of Yang-Mills instantons as was shown above, it will be interesting
to see how the topology of spacetime fabric is encoded into the
local structure of gauge fields. In particular, the representation
\eq{einstein-mfd} provides us a powerful way to prove some
inequalities about topological invariants for a closed Einstein
manifold without boundary. The Euler characteristic $\chi(M)$ and
the Hirzebruch signature $\tau(M)$ for a closed manifold $M$ are,
respectively, given by \ct{egh-report,opy}
\begin{eqnarray} \label{euler}
\chi(M) & = & \frac{1}{32 \pi^2} \int_{M} \varepsilon^{ABCD}
R_{AB} \wedge R_{CD} \\
&=& \frac{1}{2 \pi^2} \int_{M} d^4 x \sqrt{g} \Big[ \big(
f^{ab}_{(++)} \big)^2 + \big( f^{\dot{a}\dot{b}}_{(--)} \big)^2
\Big] \geq 0, \xx
\label{signature}
\tau(M) &=& \frac{1}{24 \pi^2} \int_{M} R_{AB} \wedge R_{AB} \\
&=& \frac{1}{3 \pi^2} \int_{M} d^4 x \sqrt{g} \Big[ \big(
f^{ab}_{(++)} \big)^2 - \big( f^{\dot{a}\dot{b}}_{(--)} \big)^2
\Big]. \nonumber
\end{eqnarray}
It is obvious that $\chi(M) = 0$ only if $f^{ab}_{(++)} =
f^{\dot{a}\dot{b}}_{(--)} = 0$, i.e., $M$ is flat. In addition, it
is easy to get the Hitchin-Thorpe inequality \ct{besse,egh-report}
\begin{equation} \label{hitchin-thorpe}
\chi(M) \pm \frac{3}{2} \tau(M) = \frac{1}{\pi^2} \int_{M} d^4 x \sqrt{g}
\big( f^{\dot{a}\dot{b}}_{(\pm\pm)} \big)^2 \geq 0
\end{equation}
where the equality holds if and only if $f^{ab}_{(++)} = 0$ or $
f^{\dot{a}\dot{b}}_{(--)} = 0$, i.e., $M$ is half-flat (a gravitational instanton).

The expressions \eq{euler} and \eq{signature} for topological
invariants imply that the topology of closed Einstein manifolds is
characterized by instanton and anti-instanton configurations. This
result is consistent with the Lemma in Sec. 2 stating that any
Einstein manifold is characterized by the configuration of $SU(2)_L$
instantons and $SU(2)_R$ anti-instantons. Note that the topological
invariant (instanton number $k$) of Yang-Mills instantons is given by the second Chern
character $ch_2(E)$ of the instanton bundle $E$. Therefore the Lemma
suggests that the topological invariant of Einstein manifolds should
be related to the second Chern character of instanton bundles. Let
$E_L (E_R)$ be the vector bundle for $SU(2)_L (SU(2)_R)$ Yang-Mills
instantons and $ch_2(E_L) \bigl(ch_2(E_R) \bigr)$ be the
corresponding Chern character. In order to examine the relation
between topological invariants in gravity and gauge theory, let us
consider the following decomposition
\begin{eqnarray} \label{euler-dec}
&& \chi(M) = \chi_+(M) + \chi_-(M) \equiv m \in \mathbb{Z}_{\geq 0}, \\
\label{hirzebruch-dec}
&& \tau (M) = \frac{2}{3} \bigl(\chi_+(M) - \chi_-(M) \bigr) \equiv n \in \mathbb{Z}.
\end{eqnarray}
The above decomposition can be rewritten as the form
\begin{equation}\label{+--dec}
    \chi_\pm(M) = \frac{2\chi \pm 3\tau}{4} = \frac{2m \pm 3n}{4} \geq 0
\end{equation}
where Eq. \eq{hitchin-thorpe} was used.
Note that $\chi_+(M) \bigl(\chi_-(M) \bigr)$ only depends on the $SU(2)_L (SU(2)_R)$ vector bundle.
Thus $\chi_\pm(M)$ can be a proper candidate of the Chern character $ch_2(E)$ for the instanton bundle $E$.
Indeed close inspections lead to a reasonable identification claiming that
\begin{eqnarray} \label{chern-ins}
&& k = \int_M ch_2(E_L) := \chi_+(M) =  \frac{2\chi + 3\tau}{4} = \frac{2m + 3n}{4}, \\
\label{chern-ains}
&& k = \int_M ch_2(E_R) := - \chi_-(M) =  \frac{-2\chi + 3\tau}{4} = \frac{-2m + 3n}{4},
\end{eqnarray}
where the choice of the sign in Eq. \eq{chern-ains} is to agree with
the fact that $E_R$ is the vector bundle of $SU(2)_R$
anti-instantons. Since the instanton number $k$
of gauge bundle $E$ on a closed four-manifold should take an integer
value \cite{besse,4man-book,4man-book2}, the above result predicts a nontrivial
fact that $\chi_\pm (M)$ must be integer-valued, i.e., $\chi_\pm (M)
\in \mathbb{Z}$ if $M$ is a spin manifold. It is consistent with the
result in Ref. \cite{opy} that the Euler number $\chi(M)$ for
gravitational instantons with $2\chi \pm 3\tau = 0$ coincides with
the instanton number $k$ of $SU(2)$ gauge fields. Now we will give a proof that the condition $\chi_\pm (M)
\in \mathbb{Z}$ or equivalently, $2 \chi \pm 3 \tau \in 4 \mathbb{Z}$, is satisfied
if a compact Einstein manifold $M$ is spin, i.e.,
$w_2(M)=0$.\footnote{\label{spin-c}For $\mathbb{C}P^2$ which has
$\chi=3$ and $\tau=1$, the proposed formulae give us $\int_M
ch_2(E_L) = \frac{9}{4}$ and $\int_M ch_2(E_R) = -
\frac{3}{4}$ and so they are not integers. The reason is that
$\mathbb{C}P^2$ is not a spin manifold. $\mathbb{C}P^2$ admits only
a generalized spin structure, the so-called $Spin^c$-structure
\cite{spin-book}. Therefore the decomposition
\eq{dec-riemann} for $\mathbb{C}P^2$ is valid only locally. This
caveat must also be applied to the self-dual gauge fields in Eqs.
\eq{cp2+} and \eq{cp2-}. The global spin property has to be taken
into account to describe a self-dual solution globally.}
First note that, for any compact almost complex manifold $M$,
\begin{equation}\label{rel-top}
    c_1(M)^2 = (2\chi \pm 3\tau)(M)
\end{equation}
where $c_1(M)$ is the first Chern class of $M$ \cite{4man-book}.
The first Chern class $c_1(M)$ in four dimensions satisfies the constraint
\begin{equation}\label{chern-spin}
c_1(M) \equiv w_2(M) \quad \mathrm{mod \; 2}.
\end{equation}
This is a general property for any almost complex manifold \cite{4man-book}.
For a spin manifold $M$, i.e., $w_2(M) = 0$, the constraint \eq{chern-spin} requires
$c_1(M) \in 2 \,\mathbb{Z}$. Then the relation \eq{rel-top} immediately leads to
the conclusion that  $\chi_\pm (M) \in \mathbb{Z}$ or equivalently, $2 \chi \pm 3 \tau \in 4 \mathbb{Z}$.

For a noncompact Einstein manifold, the topological invariants have
a complicated expression by including boundary terms
\ct{egh-report}. The boundary terms introduce an intricate mixing of
$SU(2)_L$ and $SU(2)_R$ gauge fields \ct{opy} whereas the bulk terms
are completely separated into two sectors as was shown in Eqs.
\eq{euler} and \eq{signature}. This mixing is triggered by the
reduction of the Lorentz group on the boundary; $SU(2)_L \times
SU(2)_R \to SU(2)_B$. For gravitational instantons where one of
$SU(2)$'s decouples from the theory, the Euler number $\chi(M)$ has
a nice interpretation in terms of the Chern-Simons form for an
$SU(2)$ vector bundle on the boundary \ct{opy}. For general Einstein
manifolds, we have not completely figured out the gauge theory
formulation of boundary terms so far. Nevertheless, because we are
using $SU(2)$ gauge fields as the basic variable, we believe that the
techniques developed for the corresponding Yang-Mills problem can be applied to the gravitational case too,
which is under study \ct{future}.

Since an Einstein manifold carries nontrivial topological
invariants, it explains why it is stable. Let us illustrate the
deconstruction \eq{einstein-mfd} of Einstein manifolds with some
examples; the Euclidean Schwarzschild metric
\ct{hawking-inst} and the Fubini-Study metric on $\mathbb{C}P^2$ \ct{gibb-pope}.
More examples and their topological properties will be discussed in
a companion paper \ct{future}. The Euclidean Schwarzschild metric is
not a gravitational instanton (not half-flat) though it is a Ricci-flat manifold \ct{hawking-inst}.
The metric takes the form
\begin{equation}\label{e-sbh}
    ds^2 = \Big(1 - \frac{2m}{r} \Big) d\tau^{2} + \Big(1 - \frac{2m}{r} \Big)^{-1} dr^{2}
+ r^{2} (d \theta^{2} + \sin \theta^2 d\phi^2).
\end{equation}
It is easy to read off the nonvanishing coefficients in Eq.
\eq{einstein-mfd} from Eq. (4.60) in Ref. \ct{opy}:
\begin{eqnarray} \label{sbh-f}
&& f^{11}_{(++)} = \frac{m}{r^3}, \qquad f^{22}_{(++)} = -
\frac{m}{2r^3} = f^{33}_{(++)}, \xx && f^{11}_{(--)} =
\frac{m}{r^3}, \qquad f^{22}_{(--)} = - \frac{m}{2r^3} =
f^{33}_{(--)}.
\end{eqnarray}
One can easily verify that the metric \eq{e-sbh} is Ricci-flat,
i.e., obeys the condition \eq{cond-ricci}.

The result \eq{sbh-f} plainly shows us that the Euclidean
Schwarzschild solution \eq{e-sbh} is the sum of an $SU(2)_L$ instanton
and an $SU(2)_R$ anti-instanton. One can show \ct{opy} that the Euler
number $\chi(M) = 1 + 1 =2 $ gets the equal contribution from the
instanton and the anti-instanton where boundary terms identically
vanish while the signature $\tau(M) = 0 - 0 =0$ is zero for both
sectors because the bulk contributions are precisely canceled by the
$\eta$-function defined by a signature operator on the boundary
\ct{egh-report}. Therefore, we checked the lemma that a Ricci-flat
four-manifold always arises as the sum of $SU(2)_L$ instantons and
$SU(2)_R$ anti-instantons and so the Ricci-flat manifold should be
stable at least perturbatively. This property is also true for an
Einstein manifold as will be examined below.

The Fubini-Study metric on $\mathbb{C}P^2$ describes a compact
K\"ahler and conformally self-dual manifold and is given by
\ct{gibb-pope}
\begin{equation}\label{cp2}
ds^2 = \frac{r^2}{4(1+ \frac{\Lambda r^2}{6})} (\sigma_1^2 +
\sigma_2^2) +
\frac{\frac{r^2}{4} \sigma_3^2 + dr^2}{(1+ \frac{\Lambda r^2}{6})^2}
\end{equation}
where $\sigma^i \, (i=1,2,3)$ are left-invariant 1-forms on the
manifold of the group $SU(2) \cong \mathbb{S}^3$ satisfying the
exterior algebra $d \sigma^i + \frac{1}{2}
\varepsilon^{ijk} \sigma^j \wedge \sigma^k = 0$.
It is straightforward to calculate the coefficients in Eq.
\eq{einstein-mfd} from the metric \eq{cp2}. The nonvanishing
coefficients are given by
\begin{equation}\label{cp2-f}
f^{33}_{(++)} = \frac{\Lambda}{2}, \qquad f^{11}_{(--)} =
f^{22}_{(--)} = f^{33}_{(--)} = \frac{\Lambda}{6}.
\end{equation}
It is clear that the metric \eq{cp2} is conformally self-dual, i.e.,
$ g^{\dot{a}\dot{b}}_{(--)} = 0$ in Eq. \eq{weyl-coeff}. One can
immediately see that the instantons and anti-instantons contribute a
ratio of three to one to both $\chi(M)$ and $\tau(M)$. Actually we get
$\chi(M) = \frac{9}{4} + \frac{3}{4} = 3$ and $\tau(M) = \frac{3}{2} -
\frac{1}{2} = 1$ \ct{egh-report}.

Note that the Euler characteristic $\chi(M)$ and the Hirzebruch
signature $\tau(M)$ are two topological invariants associated with
the Atiyah-Patodi-Singer index theorem  for an elliptic complex in
four dimensions \ct{besse,egh-report} and thus they take integer
values. In particular, as was shown in \eq{euler}, the Euler
characteristic $\chi(M)$ for any Einstein manifold $M$ takes a
positive integer unless $M$ is flat (which is true even for a
noncompact manifold). Furthermore, the inequality \eq{hitchin-thorpe} implies that there exist obstructions
to the existence of Einstein metrics in four dimensions.
For example, if $\chi(M) < \frac{3}{2} |\tau(M)|$, then $M$ does not admits any Einstein metric \ct{besse}.
Moreover, this topological consideration enhances the reason why an
Einstein manifold should be stable, at least, perturbatively.
Suppose that $M$ is an Einstein manifold such that it admits a
metric $g$ obeying \eq{einstein-equation}. Given such a metric $g$,
one can continuously perturb it to a new metric $g + \delta g$. But
the metric perturbation $g + \delta g$ cannot change the Euler
characteristic $\chi(M)$ because $\chi(M)$ is not changed by
a continuous deformation. Hence the new metric $g + \delta g$ has to
describe the same Einstein manifold as before, which means that the
continuous deformations $\delta g$ correspond to zero modes or take
values in the moduli space of Einstein metrics.
This means that the metric variation $\delta g$ with respect to the given Einstein metric $g$
has no negative eigenvalue, which is consistent with the well-known fact \cite{besse}
that the Einstein equations \eq{einstein-equation} reduce to the system of
elliptic differential operators for the metric $g$ under a suitable gauge choice,
e.g., the harmonic coordinate gauge.

\section{Einstein manifolds with a matter coupling}

Our formalism can be fruitfully applied to the deformation theory of
Einstein spaces. First of all, it will be interesting to see how the
energy-momentum tensor $T_{AB}$ of matter fields in the Einstein
equation
\begin{equation}\label{einstein-eq}
G_{AB} + \Lambda \delta_{AB} = 8 \pi G T_{AB}
\end{equation}
deforms the structure of an Einstein manifold described by Eq.
\eq{einstein-mfd}. To be specific, consider the Einstein-Yang-Mills
theory where the energy-momentum tensor of Yang-Mills gauge fields
is given by
\begin{equation}\label{ym-em-tensor}
T_{AB} = \frac{2}{g^2_{YM}} \mathrm{Tr} \Big(F_{AC} F_{BC} -
\frac{1}{4} \delta_{AB} F_{CD} F^{CD} \Big).
\end{equation}
Since the Yang-Mills field strengths $F_{AB}$ are two-forms taking
values in the adjoint representation of gauge group $G$, the Hodge decomposition \eq{2-form-dec} can
be applied to them like Eq. \eq{dec-su2l} or \eq{dec-su2r} leading to the result
\begin{equation}\label{dec-ym}
    F_{AB} \equiv f^{a}_{(+)}\eta^a_{AB} + f^{\dot{a}}_{(-)}
\overline{\eta}^{\dot{a}}_{AB}.
\end{equation}
It is then straightforward to calculate the energy-momentum tensor
\eq{ym-em-tensor} which is given by
\begin{equation}\label{dec-em-tensor}
T_{AB} = \frac{4}{g^2_{YM}} \mathrm{Tr} \big( f^{a}_{(+)}
f^{\dot{a}}_{(-)} \big)
\eta^a_{AC} \overline{\eta}^{\dot{a}}_{BC}.
\end{equation}
Substituting Eqs. \eq{dec-einstein} and \eq{dec-em-tensor} into the
Einstein equation \eq{einstein-eq} leads to the deformed relation
\begin{eqnarray}\label{def-einstein}
  && f^{ab}_{(++)} \delta^{ab} = f^{\dot{a}\dot{b}}_{(--)} \delta^{\dot{a}\dot{b}}
  = \frac{\Lambda}{2}, \xx
 && f^{a\dot{a}}_{(+-)} = \frac{16 \pi G}{g^2_{YM}}
  \mathrm{Tr} \big( f^{a}_{(+)} f^{\dot{a}}_{(-)} \big),
\end{eqnarray}
that might be compared with Eq. \eq{cond-einstein}.

The Einstein equations written in the form \eq{def-einstein} show us
a crystal-clear picture how (non-)Abelian gauge fields deform the
structure of the Einstein manifold. They introduce a mixing of
$SU(2)_L$ and $SU(2)_R$ sectors without disturbing the conformal
structure given by Eq. \eq{f-weyl} and the instanton structure
described by Eq. \eq{einstein-mfd}. This will not be the case for
other fields such as scalar and Dirac fields, as was shown in
\ct{loy1}.

An interesting but well-known point is that (anti-)self-dual
Yang-Mills fields satisfying the following equation
\begin{equation}\label{ym-sdeq}
    F_{AB} = \pm \frac{1}{2} {\varepsilon_{AB}}^{CD} F_{CD}
\end{equation}
do not affect the Einstein structure of a manifold because
$f^{\dot{a}}_{(-)} = 0$ in Eq. \eq{dec-ym} for Yang-Mills instantons
or $f^{a}_{(+)} = 0$ for anti-instantons. This is, of
course, due to the fact that the energy-momentum tensor
\eq{ym-em-tensor} identically vanishes for Yang-Mills instantons
obeying the self-duality equation \eq{ym-sdeq} \cite{opy}. Therefore
the Einstein structure is infinitely degenerate in the sense that
one can add any number of Yang-Mills instantons without spoiling the
Einstein condition of a four-manifold.

\section{Discussion}

It is a textbook statement \ct{big-book} that gravity can be
formulated as a gauge theory of local Lorentz symmetry. Nevertheless
a thorough gauge theory formulation of gravity directly reveals the
topological aspects of Einstein manifolds \ct{opy}. Indeed, as we outlined in the Introduction,
the proof of the Lemma in Sec. 2 is mainly based on the isomorphism
between the Clifford algebra and the exterior algebra.
Hence the lemma provides a completely new perspective about the
Einstein equations and Einstein manifolds. In addition, it raises a
sobering quantization issue of Einstein manifolds. The caveat is
that an Einstein manifold consists of Yang-Mills instantons. The
conventional perturbative path integral by the linearization of a
metric, $g_{MN} = \delta_{MN} + h_{MN}$, does not capture the
nontrivial topology of a vacuum Einstein manifold. In general, a
perturbative calculation around a generic background
$\overline{g}_{MN}$ will be involved with the instanton calculus
because the vacuum manifold described by the metric
$\overline{g}_{MN}$ is a configuration of Yang-Mills instantons.
Furthermore we expect that local fluctuations of spacetime geometry
in quantum gravity, the so-called quantum foams, can accompany the
quantum fluctuations of topology too and change a global structure of
spacetime fabric \ct{q-foam}. Thus it is necessary to quantize even
the vacuum geometry itself near the Planck scale. Then the problem
is how to quantize Einstein manifolds or equivalently Yang-Mills
instantons. It may be imperative to go beyond the routine approach
of quantum gravity.

Our result \eq{ym-instanton-eq} can be applied to find an $SU(2)$
Yang-Mills instanton on a general Einstein manifold which
generalizes the result in Refs. \ct{charap-duff,charap-duff2} for Ricci-flat manifolds.
Given an Einstein metric $g$, one can calculate the spin connections
and Riemann curvature tensors of the Einstein metric $g$. Our lemma
then says that the self-dual and anti-self-dual spin connections in
Eq. \eq{spin-sd-asd} are automatically $SU(2)$ (anti-)self-dual
connections obeying Eq. \eq{ym-instanton-eq} defined on the Einstein
manifold whose metric is given by $g$. It may be more transparent by
rewriting Eq. \eq{ym-instanton-eq} as the form \ct{opy}
\begin{equation}\label{su2-lo-instanton}
     F^{(\pm)}_{MN} = \pm \frac{1}{2}\frac{\varepsilon^{RSPQ}}{\sqrt{g}}
     g_{MR}g_{NS}  F^{(\pm)}_{PQ}
\end{equation}
where $\sqrt{g} = \det E_M^A$ and $\varepsilon^{RSPQ}$ is the metric
independent Levi-Civita symbol with $\varepsilon^{1234} = 1$.

Let us illustrate with the Fubini-Study metric \eq{cp2} on
$\mathbb{C}P^2$ that, whenever an Einstein metric is given, it is
possible to find an $SU(2)$ Yang-Mills instanton on the
Einstein manifold. Using the torsion free condition, $T^A = dE^A +
{\omega^A}_B \wedge E^B = 0$, it is easy to get the spin connections
and so $SU(2)$ gauge fields in Eq. \eq{spin-sd-asd} for the metric \eq{cp2}:
\begin{eqnarray} \la{cp2+}
&& A^{(+)1} =  A^{(+)2} = 0, \quad  A^{(+)3} = - \frac{\Lambda r}{4} E^3,  \\
\la{cp2-}
&& A^{(-)1} = - \frac{1}{2r} E^1, \;\; A^{(-)2} = - \frac{1}{2r}
E^2, \;\; A^{(-)3} = - \frac{1}{2r} (1+ f) E^3,
\end{eqnarray}
where
\begin{equation}\label{cp2-tetrad}
    E^1 = \frac{r}{2\sqrt{f}} \sigma^1, \;\;  E^2 = \frac{r}{2\sqrt{f}}
    \sigma^2, \;\;  E^3 = \frac{r}{2f} \sigma^3
\end{equation}
with $f(r) = 1 + \frac{\Lambda r^2}{6}$. It is straightforward to
check that the self-dual gauge fields in Eq. \eq{cp2+} and the
anti-self-dual gauge fields in Eq. \eq{cp2-} separately obey the
self-duality equations in Eq. \eq{su2-lo-instanton} (with $+$-sign and
$-$-sign, respectively), as was already verified in Eq. \eq{cp2-f}.
Hence the Fubini-Study metric \eq{cp2} can be used in this way to
find $SU(2)$ Yang-Mills instantons on $\mathbb{C}P^2$.
But, as we pointed out in the footnote \ref{spin-c}, the $SU(2)$ gauge fields in Eqs. \eq{cp2+}
and \eq{cp2-} should be understood only locally. This is due to the fact that
$\mathbb{C}P^2$ is not a spin manifold, hence the spin connection
cannot be split globally without sign ambiguity. But the sign ambiguity does not affect
the instanton structure for $\mathbb{C}P^2$ described by Eq. \eq{cp2-f}.

It should be interesting to investigate a generalization of gauge
theory formulation of Einstein gravity to other dimensions. Such a
generalization to six dimensions was already considered using the
Lie algebra isomorphism between $SO(6)$ Lorentz algebra and $SU(4)$
Lie algebra \ct{yayu-6}. Of course, three and five dimensions can
also be invited to the gauge theory formulation. In three
dimensions, spin connections and Lorentz generators can be
identified with $SU(2)$ gauge fields and Lie algebra generators,
respectively, as $\omega_{AB} \equiv \varepsilon_{ABC} A^C$ and
$J^{AB} \equiv \varepsilon^{ABC} T^C$, and $T^A$ obey the commutation
relation $[T^A, T^B] = - \varepsilon^{ABC} T^C$. Then one can show
that $R_{MNAB} \equiv \varepsilon_{ABC} F_{MN}^C$ where $F_{MN} =
\partial_M A_N - \partial_N A_M + [A_M, A_N]$ with $A_M = A_M^A
T^A$. After some work, it can be shown that an Einstein manifold
satisfying the condition $R_{AB} = \Lambda
\delta_{AB}$ corresponds to $F_{AB}^C = \frac{1}{2} \varepsilon_{ABC}
\Lambda$. In particular, a Ricci-flat manifold with $\Lambda = 0$ is described
by $SU(2)$ flat connections, i.e., $F_{AB} = 0$.

In five dimensions, the Lorentz group is $SO(5) = Sp(2)$ which is a
simple Lie group. Therefore, five-dimensional gravity can be recast
in the form of $Sp(2)$ Yang-Mills gauge theory. An interesting
problem is to understand how $SU(2)$ Yang-Mills instantons and
anti-instantons for four-dimensional Einstein manifolds can be
embedded together into the simple group $SO(5) = Sp(2)$ and what is
a corresponding gauge theory object for five-dimensional Einstein
manifolds. In particular, if we consider a Kaluza-Klein
compactification along the fifth direction, the five-dimensional
metric will take the form
\begin{equation}\label{kk-metric}
    ds^2 = e^{\phi/\sqrt{3}} \Big( g_{MN} dx^M dx^N + e^{- \sqrt{3} \phi} (dx^5 + A_M dx^M)^2
    \Big).
\end{equation}
It is well-known that the resulting five-dimensional gravity reduces
to Einstein-Maxwell-dilaton theory in four dimensions. It is then
interesting to see how this Einstein-Maxwell-dilaton theory is
embedded in $Sp(2)$ Yang-Mills gauge theory in five dimensions. Our
gauge theory formulation here can be plainly applied to the
four-dimensional gravity part coupling to $U(1)$ gauge fields and a
dilaton. A detailed analysis for the gauge theory formulation of
three- and five-dimensional gravity will be reported elsewhere.

\section*{Acknowledgments} We thank Jungjai Lee and Chanyong Park for helpful
discussions. This work was supported by the National Research Foundation of Korea (NRF) grant
funded by the Korea government (MOE) (No. 2011-0010597).
The work of H.S. Yang was also supported by the RP-Grant 2010 of Ewha Womans University.

\end{document}